The formation of NV centers in diamond: A theoretical study based on calculated transitions and migration of nitrogen and vacancy related defects.

Peter Deák,[*,1] Bálint Aradi,[1] Moloud Kaviani,[1] Thomas Frauenheim[1]

and

Adam Gali[2,3]

[1] Bremen Center for Computational Materials Science, University of Bremen, PoB 330440, D-28334 Bremen, Germany

[2] Wigner Research Center for Physics, Hungarian Academy of Sciences, PoB 49, H-1525, Budapest, Hungary

[3] Department of Atomic Physics, Budapest University of Technology and Economics, Budafoki út 8., H-1111, Budapest, Hungary

ABSTRACT

Formation and excitation energies as well charge transition levels are determined for the substitutional nitrogen ($N_s$), the vacancy (V), and related point defects (NV, NVH, $N_2$, $N_2V$ and $V_2$) by screened non-local hybrid density functional supercell plane wave calculations in bulk diamond. In addition, the activation energy for V and NV diffusion is calculated. We find good agreement between theory and experiment for the previously well-established data, and predict missing ones. Based on the calculated properties of these defects, the formation of the negatively charged nitrogen-vacancy center is studied, because it is a prominent candidate for application in quantum information processing and for nanosensors. Our results indicate that NV defects are predominantly created directly by irradiation, while simultaneously produced vacancies will form $V_2$ pairs during post-irradiation annealing. Divacancies may pin the Fermi-level making the NV defects neutral.

I. INTRODUCTION

Diamond has point defects which could act as quantum bits in quantum information processing at room temperature.[1,2] Particularly, the nitrogen-vacancy (NV) defect[3,4,5] is a prominent candidate, and can also be used as a nanosensor to detect magnetic[6,7,8,9,10,11] and electric[12] fields, temperature,[13,14,15] or chemical changes on the surface.[16] All these applications rely on the negative charge state of NV which, however, can be often found also in the neutral charge state in bulk diamond.[4,17] In addition, the negative center NV(-) may temporarily or permanently lose its charge state during optical excitation.[18,19,20,21] The photo-stability may be improved by post-annealing treatments.[22]

[*] Corresponding author: deak@bccms.uni-bremen.de

NV(-) can be routinely found in natural Type Ib diamonds but, generally, the concentration of NV-defects in natural or as-grown synthetic diamond is too low for applications.[23] The concentration of NV-defects can be substantially increased by creating vacancies in nitrogen-doped diamond by irradiation with energetic neutrons, electrons or ions,[24,25] followed by annealing above ~ 600 °C, where vacancies become mobile.[4,26] According to the present consensus in the literature, mobile vacancies are then trapped by substitutional nitrogen ($N_s$), forming NV-centers.[27,28,29,30,31]

Understanding the formation of NV defects in as-grown or irradiated diamond samples requires the accurate knowledge about the formation energy of the isolated constituents: $N_s$ and the vacancy (V), and also about competing defect complexes. The mobility of the species and the energy of complex formation may depend on the charge states, thus it is highly critical to determine the charge transition levels of these defects across the band gap. Since diamond is a wide gap material, it is extremely difficult (or sometimes even impossible) to determine deep adiabatic (thermal) charge transition levels by traditional techniques such as deep level transient spectroscopy (DLTS). The vertical ionization energies may be obtained by optical excitation of the samples but it is not trivial to interpret the signals from these experiments. However, recent advances in density functional theory have made it possible to calculate transition levels with very good accuracy.[32]

In this paper we apply advanced density functional theory calculations to determine the formation and excitation energies, the charge transition levels and diffusion activation energies for nitrogen and vacancy related defects in diamond. We have been able to reproduce the known data with good accuracy, and predict the missing ones, which are needed to study the complex formation of these defects in as-grown as well as irradiated diamond samples. We note that the effect of extended defects (the surface, grain boundaries, voids or aggregates) on the formation and charge state of NV is beyond the scope of this paper. We believe that the first step towards following and understanding the atomistic processes of NV creation should inevitable be taken in bulk diamond, considering the most simple and relevant reaction paths only. Here, we focus our study particularly on the formation of small complexes, such as the divacancy ($V_2$), the pair of substitutional nitrogen atoms ($N_2$), the NV as well as the $N_2V$ and NVH centers, from isolated constituents, considering all possible charge states of these defects.

We find that the concentration of NV in as-grown diamond is always at least three orders of magnitude smaller than that of $N_s$, due to the low equilibrium concentration of vacancies. The calculated reaction energies between $N_s$ and V defects indicate that the concentration of NV will not be higher even if a non-equilibrium excess of vacancies are provided, due to the preference for $V_2$ over NV formation. We show, however, that NV formation can be expected to dominate over V formation during irradiation. We also find that $V_2$ defects crucially influence the charge state of NV, and having the latter predominantly in the negative charge state requires the reduction of the divacancy concentration.

The paper is organized as follows. In Section II the methodology is described in detail. In Section III we provide the results. We analyze each point defect in separate Sections

III.A-G where we compare our calculations with existing experimental or theoretical data from previous work. We discuss the formation and charge state of NV in Section III.H. Finally, we briefly summarize the results in Section IV.

II. METHODS

Defect calculations in solids are almost always carried out by applying two basic approximations: *i)* the adiabatic principle, i.e. the separation of the electron problem from that of the lattice vibrations and *ii)* the one-electron approximation, which expresses either the wave function (in Hartree-Fock theory) or the density (in the Kohn-Sham theory) of the many-electron system in terms of independent single-particle states. The neutral vacancy in diamond is the schoolbook example for the failure of both of these approximations. Strong electron-phonon coupling gives rise to a dynamic Jahn-Teller effect, obliterating in room-temperature measurements the static Jahn-Teller distortion predicted by theoretical calculations at 0K, and the degenerate ground state cannot be described with just one single-particle configuration. Still, the system sizes necessary to model vacancy-related defects in the solute limit are just too big for abandoning these approximations. Therefore, they will still be used in this study too, in the hope that in calculated *energy differences* the lack of many-body effects and electron-phonon coupling causes errors of 0.1-0.2 eV at most due to error compensation. As we will show, comparison of our results to experimental data supports this expectation.

Nitrogen and vacancy related defects have been investigated very thoroughly earlier using density functional theory (DFT) within the local density or the generalized gradient approximation (LDA and GGA, respectively), and by semi-empirical methods.[25,33,34,35,36,37,38] While these studies have revealed the basic configurations of the relevant defects, calculated gap levels and optical transitions were impaired even in *ab initio* calculations by the electron self-interaction error involved with the standard approximations of DFT. Precise calculation of these data are very important for defect identification, but the correct reproduction of the defect levels is also crucial for calculating relative energies of different configurations, and for the activation energy of diffusion.[39] The present calculations have been carried out in the framework of the generalized Kohn-Sham theory,[40] by using the screened hybrid functional HSE06 of Heyd, Ernzerhof and Scuseria with the original parameters (0.2 Å$^{-1}$ for screening and 25% mixing).[41] Previously we have shown[32] that defect levels calculated with this method in Group-IV semiconductors fulfill the generalized Koopmans' theorem,[42] i.e., the total energy is a linear function of the fractional occupation number. Due to the error compensation between the Hartree-Fock and GGA exchange (which would lead, if applied purely, to concave or convex total energies, respectively), HSE06 in diamond happens to be nearly free of the electron self-interaction error, and is capable of providing defect-levels and defect-related electronic transitions within ~0.1 eV to experiment.[32,43]

We have used the Vienna *ab-initio* simulation package VASP5.2.12 with the projector augmented wave method (applying projectors originally supplied to the 5.2 version).[44] To avoid size effects as much as possible, a 512-atom supercell was used in the Γ-approximation for defect studies. Parameters for the supercell calculations were established first by using the GGA exchange of Perdew, Burke and Ernzerhof (PBE)[45] in bulk calculations on the primitive cell with a 8×8×8 Monkhorst-Pack (MP) set for Brillouin-zone sampling.[46] (Increasing the MP set to 12×12×12 has changed the total energy by < 0.002 eV.) Constant volume relaxations using a cutoff of 370 (740) eV in the plane-wave expansion for the wave function (charge density) resulted in an equilibrium lattice parameter of $a_{PBE}$ = 3.570 Å. Increasing the cutoff to 420 (840) eV has changed the lattice constant by only 0.003 Å. Therefore, considering the demands of the supercell calculations, the lower cut-off was selected. An HSE06 calculation with the 8×8×8 MP set and 370(740) eV cutoff resulted in the lattice constant $a_{HSE}$ = 3.545 Å, the bulk modulus $B_0$ = 425 Å and the indirect band gap of $E_g$ = 5.34 eV, in good agreement[47] with the experimental values of $a$ = 3.567 Å, $B_0$ = 443 GPa and $E_g$ = 5.48 eV (see, e.g., Ref.[39]). Due to the different choice of the basis, the HSE06 values presented here differ somewhat from those in Refs. [32,39,43], but tests on the NV(-) center have shown that the higher cutoff would cause only very small differences in the equilibrium geometry of that defect too.

Defects in the supercell were allowed to relax in constant volume till the forces were below 0.01 eV/Å. Diffusion activation energies were determined by the nudged elastic band method (NEB).[48] For comparison of different defect configurations and charge states, the electrostatic potential alignment and the charge correction scheme of Lany and Zunger was applied.[49,50] Recently, this scheme was found to work best for defects with medium localization.[51]

Experimental diffusion studies in diamond are performed at high temperatures (800-2200K), so approximating the free energy of diffusion activation with the energy is quite inaccurate. The strongest temperature dependent contribution to the free energy in diamond comes from the vibrations. The vibration energy and entropy have been estimated by density-functional based tight binding (DFTB)[52] calculations, as described earlier for the vacancy in silicon carbide,[53] using the DFTB+ code.[54]

III. RESULTS AND DISCUSSION

A. *Substitutional nitrogen* ($N_s$) is the most prominent defect of Type Ib natural and N-doped CVD (chemical vapor deposited) diamond, and it has been thoroughly studied experimentally. It is stable up to high temperatures, with a diffusion activation energy of 5.0±0.3 eV (as measured between 1700-2100 °C at a pressure of 7 GPa).[55] In another high-temperature/high-pressure experiment, a lower barrier of 2.6 eV was found,[56] presumably due to the assistance of intrinsic defects generated by pressure effects.[55] Theoretical calculations (at 0K) find an activation energy of 6.3 eV for the direct exchange of $N_s$ with a neighbor C atom,[25] while the rate limiting step for vacancy assisted diffusion was found to be the jump of $N_s$ into a next neighbor vacancy (V), with a calculated barrier of about 4.8 eV.[33,34] The optical signature of the $N_s$ defect is well known. In UV absorption the *A* band at 3.3 eV and the *B* band at 3.9 eV were assigned to vertical transitions from the $A_1$ ground state of the defect to effective-mass-like $A_1$ and E excited states, respectively.[57,58] From the thermal activation of the conductivity, the adiabatic (+/0) charge transition level of $N_s$ was found to be with respect to the conduction band edge ($E_C$) at $E_C$ – 1.7 eV.[59] A negatively charged state of $N_s$ has been predicted theoretically and confirmed experimentally.[60,61] It has been suggested that the zero phonon line (ZPL) measured at 4.059 eV in Type Ib diamonds is associated with the formation of $N_s(-)$, by populating $N_s(0)$ with an additional electron from the valence band. The absorption band at 4.6 eV was assigned to the corresponding vertical transition.[57,62] LDA calculations find $N_s$ to have $C_{3v}$ symmetry, with the distance of $N_s$ to the nearest C-neighbor along the trigonal axis being ~28% longer than the C–C bonds.[33] With the application of the marker method – to correct for the deficiencies of LDA and the cluster model – the vertical (adiabatic) charge transition levels were estimated to be at $E_C$ – 2.9 eV ($E_C$ -1.5 eV) for the (+/0) and at $E_V$ + 4.7 eV ($E_V$ + 4.4 eV) for the (0/-) transition, respectively.[35,60]

**Table 1**. Comparison of the vertical and adiabatic charge transition levels, calculated in this work by the HSE06 functional, with experiment. Numbers in parentheses are the estimates with the marker method, based on LDA calculations,[60] and the results of another HSE06 calculation in a smaller 64-atom supercell with a 400 eV cutoff.[63] Donor levels are given with respect to the conduction band edge $E_C$, acceptor levels with respect to the valence band edge $E_V$ (in eV).

| Defect | Charge transition level | Vertical | | Adiabatic | | |
|---|---|---|---|---|---|---|
| | | HSE06 (LDA[60]) | Exptl. | HSE06 (LDA;[60] HSE06[63]) | | Exptl. |
| $N_s$ | (+/0) | $E_C$ - 3.1 (2.9) | $E_C$– 3.3[a)] | $E_C$ - 1.8 (1.5; | 1.8) | $E_C$– 1.7[b)] |
| | (0/-) | $E_V$ + 4.9 (4.7) | | $E_V$ + 4.6 (4.4; | 4.5) | |
| V | (2+/+) | $E_C$– 5.0 | | $E_C$– 4.9 | | |
| | (+/0) | $E_C$– 4.5 | | $E_C$– 4.4 (—; | 4.4) | $E_C$– 4.3[c)] |
| | (0/-) | $E_V$ + 2.1 | | $E_V$ + 2.0 (—; | 1.9) | |
| | (-/2-) | $E_V$ + 4.8 | | $E_V$ + 4.9 | | |
| NV | (+/0) | $E_C$– 4.6 | | $E_C$– 4.4 (—; | 4.7) | |
| | (0/-) | $E_V$ + 2.7 | | $E_V$ + 2.7 (—; | 2.8) | |
| | (-/2-) | $E_V$ + 4.9 | | $E_V$ + 4.9 | | |
| $N_2$ | (+/0) | $E_C$ – 4.4 | | $E_C$ – 4.0 | | $E_C$ – 4.0[d)] |
| $N_2V$ | (+/0) | $E_C$ – 4.8 | | $E_C$ – 4.7 | | |
| | (0/-) | $E_V$ + 3.3 | | $E_V$ + 3.2 | | |
| $V_2$ | (+/0) | $E_C$ – 4.3 | | $E_C$ – 4.3 | | |
| | (0/-) | $E_V$ + 2.4 | | $E_V$ + 2.3 | | |
| | (0/2-) | $E_V$ + 3.2 | | $E_V$ + 3.2 | | |
| NVH | (+/0) | $E_C$ – 4.9 | | $E_C$ – 4.5 | | |
| | (0/-) | $E_V$ + 2.6 | | $E_V$ + 2.4 | | $E_V$ + 2.4[e)] |
| | (-/2-) | $E_V$ + 4.6 | | $E_V$ + 4.4 | | |

a) Since the excited effective-mass-like states in diamond are within 0.1 eV of the band edges, within the accuracy of the calculations the vertical ionization energy of N can be compared to the observed A band of the optical absorption spectrum [57].
b) Thermal activation energy of conductivity: Ref.[57]
c) DLTS: Refs. [64,65]
d) Photoconductivity [73]
e) Absorption [23]

Our HSE06 calculation reproduces the $C_{3v}$ symmetry (with one N–C distance elongated by 32%), but also *all* the experimentally observed electronic transitions with an accuracy better than 0.2 eV and without any *a posteriori* correction. Table I. shows the vertical and adiabatic charge transition levels. To check the creation mechanism of $N_s(-)$, we have attempted to calculate an exciton with the electron trapped in the gap level of $N_s$, but the hole also got localized into a defect-related state above the valence band (VB) edge. The vertical excitation energy and the corresponding ZPL for creating such an excited state of $N_s$ are in excellent agreement with the experimental values (see Table 2). Since nitrogen diffusion without the assistance of vacancies only occurs at temperatures and pressures irrelevant for the application of the NV center, we have not attempted to calculate the energy barrier for direct exchange of $N_s$ with a neighbor C atom. Vacancy assisted nitrogen diffusion will be considered in the section about the NV defect.

**Table 2**. Intra-defect vertical transitions and the corresponding ZPL in eV.

| Defect | Transition | Vertical | | ZPL | |
|---|---|---|---|---|---|
| | | HSE06 | Exptl. | HSE06 | Exptl. |
| $N_s(0)$ | $^1A_1 \to {}^1A_1$ | 4.6 | 4.6[a] | 4.1 | 4.1[a] |
| $V(-)$ | $^4A_2 \to {}^4T_1$ | | | 3.3 | 3.3[b] |
| $NV(-)$ | $^3A_2 \to {}^3E$ | 2.3 | 2.2 | 2.0 | 2.0[c] |
| $N_2$ | $^1A_{1g} \to {}^1A_u$ | 4.0 | | 3.6 | 3.8[d] |
| $^1N_2V(0)$ | | 2.8 | | 2.7 | 2.5[e] |
| $^3N_2V(0)$ | | 2.7 | | 2.6 | 2.5[e] |

a) Refs.[57,62]
b) see. e.g. Ref.[67]
c) Ref.[4]
d) Ref.[72]
e) Ref.[66]

B. *The single vacancy* (V) is the origin of numerous bands in the optical spectra of diamond. The ZPL of the GR1 band at 1.67 eV, and of the ND1 band at 3.15 eV, are assigned to the excitation of the neutral and the negative vacancy, V(0) and V(-), respectively (see. e.g. Ref.[67]). Based on DLTS studies, the adiabatic (+/0) charge transition level was suggested to be at 1.25 eV or 1.13 eV above the valence band (Refs.[64] and [65], respectively), corresponding to about $E_C - 4.3$ eV. The (0/-) level is expected to be around midgap,[67] which would be hard to detect directly. V(0) is mobile between 600 and 800 °C with an activation energy of $2.3 \pm 0.3$ eV, while V(-) is not: the vacancy probably undergoes a charge transition before diffusing.[67]

Theoretically, the unrelaxed vacancy gives rise to a non-degenerate $a_1$ and a triply degenerate $t_2$ single-particle defect state, the latter higher in energy. An $a_1(\uparrow\downarrow)$, $t_2(\uparrow\downarrow:0:0)$ singlet configuration for V(0) is Jahn-Teller-unstable. LDA calculations result in a $D_{2d}$ distortion, with the axial displacement of the first neighbors (parallel to the main symmetry axis) much larger than the radial one.[36] This splits the triply degenerate single-particle state $t_2(\uparrow\downarrow:0:0)$ into $b_2(\uparrow\downarrow) + e(0:0)$. As mentioned in the Methods section, the experimental results on the vacancy can be analyzed in terms of many-body states in $T_d$ symmetry.[68] The singlet $^1E$ ground state of V(0) cannot be described by just one single-particle configuration. The single-particle configuration $a_1(\uparrow\downarrow)$, $b_2(\uparrow\downarrow) + e(0:0)$, obtained from LDA calculations, is a weighted sum of the $^1E$ and the $^1A_1$ many-body states. Performing GGA calculations (with the PBE exchange functional), we could reproduce the $D_{2d}$ state described above for V(0), but we also found a metastable state, 0.24 eV higher in energy, where the axial distortion is smaller than the radial, and the splitting of the $t_2$ single-particle state gives rise to a doubly degenerate *e* level *lower in energy* than the non-degenerate $b_2$ state. In the HSE06 calculation this latter situation turns out to be the ground state. A singlet $a_1(\uparrow\downarrow)$, $e(\uparrow:\downarrow) + b_2(0)$ occupation is, in principle, Jahn-Teller unstable, but upon relaxation the geometry very nearly preserves the $D_{2d}$ symmetry, with the two spin-orbitals, which belong to one given level (after the splitting of $e(\uparrow:\downarrow)$ into $b_1(\uparrow)+b_2(\downarrow)$ states), having orthogonal mirror planes. The radial distortion is 12%, the axial 5%. This singlet configuration is 0.18 eV lower in energy than a triplet $a_1(\uparrow\downarrow)$, $e(\uparrow:\uparrow) + b_2(0)$, with 11 % radial and 8% axial distortion. On the one hand, this might very well be an artifact of the hybrid functional due to the overestimated strong splitting of the *e*-level.[69] On the other, the triplet $a_1(\uparrow\downarrow)$, $e(\uparrow:\uparrow) + b_2(0)$ single-particle configuration is one of those degenerate ones which make up the $^3T_1$ excited state of V(0). The latter is known to be 0.1 eV above the singlet $^1E$ ground state.[70] Since this difference is within the error bar of our calculations, we decided to use the triplet $a_1(\uparrow\downarrow)$, $e(\uparrow:\uparrow) + b_2(0)$ single-particle configuration as reference state for V(0). It is reasonable to assume that we are committing the same error when describing V(+) with a single $a_1(\uparrow\downarrow)$, $b_1(\uparrow:0) + a_1(0) + b_2(0)$ configuration in $C_{2v}$ symmetry. The resulting (+/0) charge transition level (Table 1) is indeed within 0.1 eV to the experimental observation. The (0/-) level is predicted at $E_V$ + 2.0 eV, i.e., indeed close to midgap.[67]

**Table 3**. Diffusion activation energies (in eV) from HSE06 (this work) without and with DFTB corrections for the vibrational energy and entropy at 1000K, compared to high temperature experimental data. (Numbers in parenthesis are from the LDA calculations of Ref.[34,36])

| Defect | HSE06 (LDA) | HSE06+DFTB | Exptl. |
|---|---|---|---|
| V(0) | 2.8    (2.8) | 2.6 | 2.3±0.3[a] |
| V(-) | 3.5    (2.5) |  | immobile [a] |
| NV(0) | 4.7    (4.8) | 4.5 |  |

a) Ref.[67]

The ground state of V(-) is $^4A_2$ in $T_d$ symmetry, to which only one single-particle configuration is contributing: $a_1(\uparrow\downarrow)$, $t_2(\uparrow:\uparrow:\uparrow)$, which is stable against static Jahn-Teller distortion.[36] The same is true for the $^4T_1$ excited state of V(-) which is given by the three degenerate $a_1(\uparrow)$, $t_2(\uparrow\downarrow:\uparrow:\uparrow)$ single-particle configurations. The vertical transition from the ground state to this excited state was calculated to be 3.3 eV with LDA,[36] even though a higher energy is expected than the ZPL observed at 3.3 eV. The same calculation resulted in diffusion activation energies of 2.80 eV and 2.47 eV for V(0) and V(-), respectively, with the saddle point being off the [111] direction in the $(1\bar{1}0)$ plane. While the first value is reasonably close, the second contradicts the experiments which indicate a much higher activation energy for V(-).[67] Our HSE calculation for V(-) results in 3.3 eV for the ZPL of the ND1 band (Table 2), in excellent agreement with experiment. Our results also indicate that the vacancy can be stable in 2+ and 2– charge states too (Table 1). In calculating the diffusion barriers by the NEB method, we have followed the route given in Ref. [36]. While our result for V(0) is identical with that of the LDA calculation (see Table 3), the HSE06 barrier for V(-) is substantially higher, giving rise to diffusivities $10^6$ times smaller than that of V(0) at 1000 K, in agreement with experiment.[67] Taking into account the energy and entropy of vibrations at 1000K, the calculated free energy of activating the diffusion of V(0) is 2.6 eV, which is within the bounds of the experimental determination, especially when considering the neglect of many body effects.

C. *The nitrogen-vacancy center* (NV). The NV(-) color center of diamond is at the focus of many experimental and theoretical papers (see e.g. Refs.[43, 71 ]). The observed vertical absorption and the ZPL of the NV(-) center can be well reproduced by HSE06 calculations. Note that the values given in Table 2 differ slightly from our previously published result,[43] due to the changes in the parameters of the calculation, but agree well with those of Ref.[63]. We have not calculated the internal transitions of the NV(0) center, because its excited state cannot be described by one single-particle configuration. There are no experimental data available on the charge transition levels. Our calculated adiabatic values for the (+/0) and (0/–) levels in Table 1 are shallower than those of Ref. [63], probably due to the use of the much larger supercell. According to our calculations, an NV(2-) charge state could in principle also exist, but donors shallower than $N_s$ would be needed to obtain them (see Fig.1). Our calculated energy barrier for the jump of the N atom into the vacancy agrees well with the LDA values of Refs. [25,34], but the free energy of activation at high temperature (Table 3) is still much higher than the 2.6 eV observed during the aggregation of dispersed $N_s$ into pairs (called *A* aggregate).[56] Although the latter happens to be close to the activation energy of V(0) diffusion, our result indicates that the aggregation process could not have been assisted by vacancies (leaving only self-interstitials as possible mediators). According to the mechanism proposed in Ref. [34] the N→V jump is also the critical step for the diffusion of the NV-center. Our barrier of 4.5 eV indicates that NV centers will remain immobile up to about 1700 °C, unless, maybe, if self-interstitials are released from larger aggregates.

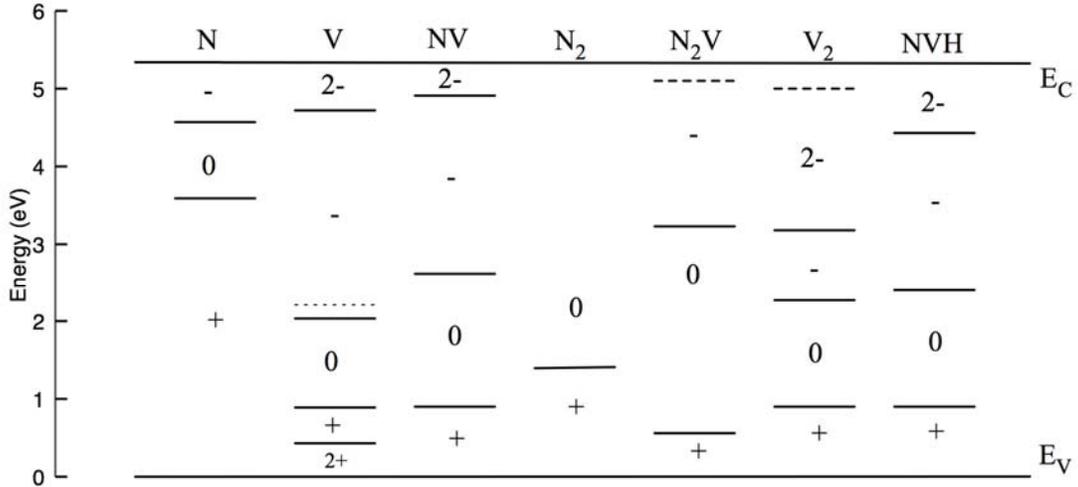

**Fig.1**. Comparison of the adiabatic charge transition levels. Dashed lines are estimates based on the vertical ionization energy, computed from the frontier orbitals. The dotted line for the single vacancy arises from an estimate for the singlet many-body ground state (0.1 eV below the first triplet many-body state used for obtaining the solid line.)

D. *The $N_2$ defect*, i.e., two first neighbor $N_s$ was identified as the "*A*-aggregate" in Type IaA diamond.[25,38,72] It is characterized by a ZPL at 3.8 eV. Such defects can be found only in natural diamonds, or after annealing above 2000K. It is assumed that the ZPL is connected to a hyper deep donor level at $E_C$-4.0 eV.[73] Since this level might influence the charge state of, and is an obvious trap for vacancies (forming the $N_2V$ center), we have calculated its electronic structure. The ground state is singlet, with just one doubly occupied level in the gap. As shown in Tables 1-2, the HSE results for the (+/0) charge transition level, as well as the ZPL of the internal excitation of the defect are in good agreement with experiment. Interestingly, the $N_2$(+) defect has no occupied level in the gap, so no other charge states than 0 and + can be expected.

E. *The $N_2V$ center*. NV centers are created by the capture of mobile V(0) at substitutional $N_s$, however, nitrogen clusters are competing traps for the vacancies. The smallest such complex is the so called *A* aggregate, ($N_2$ or $N_s$-$N_s$, see above). By the capture of a vacancy an $N_s$-V-$N_s$ complex with $C_{2v}$ symmetry is formed (i.e., the vacancy has now two nitrogen and two carbon neighbors).[25,74] This defect gives rise to the *H3* optical center, characterized by a ZPL of 2.463 eV due to a transition between the singlet $^1A_1$ and $^1B_1$ states.[75] Time-dependent measurements show also a delayed luminescence with about the same energy from a triplet state, energetically very close to $^1B_1$.[76] LDA calculations[74] predict the occupied bonding and the unoccupied antibonding state between the two carbon neighbors to the vacancy (at a distance of 2.58 Å) as having levels in the gap. The calculated vertical transition energy was found to be 0.93 eV, and the discrepancy with the observed value was attributed to the LDA approximation.

Our (spin-polarized) HSE06 calculation for this complex results in a somewhat larger distance between the carbon neighbors, 2.71 Å, and three levels in the gap. The lowest one is a doubly occupied level 0.1 eV above the valence band edge, and corresponds to a state weakly localized on the nitrogen neighbors in antibonding combination. The singlet ground state is "antiferromagnetic": the spin-up and spin-down wave functions of the next occupied gap level are localized on either one of the two carbon neighbors. The same is true for the third, unoccupied level. Obviously, such ↑ and ↓ states alone in themselves do not correspond to the $C_{2v}$ symmetry of the system, yet a relaxation without constraint preserves that symmetry. Apparently the two dangling $sp^3$ hybrids of the carbon neighbors represent a biradical state which, similarly to the four dangling bonds of the single vacancy, cannot be described with just one single-particle configuration. We have calculated the excitation energies on the assumption that picking just one such configuration for both the ground and the excited state will give a reasonable estimate of the true many-body excitations. For the vertical excitation between the carbon related occupied and unoccupied states we have obtained 0.82 eV, close to the LDA prediction. Adiabatic excitation from the lowest gap state (weakly localized to the nitrogen neighbors) to the unoccupied carbon-related state, however, gave reasonably good agreement with the experimentally observed ZPL. We have also found a metastable triplet state, 0.18 eV above the "antiferromagnetic" singlet ground state. In this case one electron is on each of the bonding and the antibonding combination of the $sp^3$ hybrids of the two carbon neighbors. Also an excited triplet state exists (with one electron promoted from the antibonding N-N state to the bonding C-C state), 0.29 eV higher than the excited singlet state described above. This leads to two recombination channels with similar

ZPLs, as indeed measured in experiment.[76] These results shows that, despite of the limitations due to the single particle approximation, the HSE06 results are correct on a semi-quantitative level. In calculating the charge transition levels, we have taken the antiferromagnetic singlet ground state as reference, and expect a similar uncertainty in the values given in Table 1 as in the case of the single vacancy.

F. *The divacancy*. Obviously, when neutral vacancies start to diffuse, divacancies may also form in various charge states. This has to be taken into account, too, when considering the equilibrium concentration of NV(-) centers. The neutral divacancy, $V_2(0)$, has signatures both in paramagnetic and optical spectra.[77,78] Analysis of the former has led to the surprising conclusion that the ground state of $V_2(0)$ is a triplet in $C_{2h}$ symmetry, instead of the intuitively expected $D_{3d}$.[79] Comer et al. interpreted this as a result of a level crossing (similar to our case for the single vacancy), due to a strong outward relaxation of those carbon pairs, which do not lie in the mirror plane.[37] While this $C_{2h}$ structure was 0.1 eV higher in energy than a $D_{3d}$ one in LDA, it turns out to be the ground state in our HSE06 calculation, being 0.07 eV lower in energy. In the HSE06 ground state, the gap contains only states derived from the $e_u$ and $e_g$ states of the ideal divacancy, of course split up in accord with the $C_{2h}$ symmetry. Here again only the sum of the two spin orbitals of the same level transform according to the irreducible representations of the $C_{2h}$ point group, but relaxation without constraint preserves the $C_{2h}$ symmetry. In addition, the lowest energy singlet and the triplet single-particle configurations have the same energy within the accuracy of the calculation. All this points to a many body ground state which cannot be well described in a single-determinant approximation. In our calculations all vertical excitations between the gap states have lower energy than the ZPL attributed to $V_2(0)$ at 2.543 eV.[77] It appears likely that an $a_1$ state (still visible in the gap under the $D_{3d}$ symmetry constraint) contributes to the excitation, however this cannot be taken into account in our one-determinant approximation. The negatively charged divacancy, $V_2(-)$, was proposed as the origin of the W29 paramagnetic center with quadruplet spin state.[80] We find a quadruplet ground state for $V_2(-)$, supporting this assignment. As shown in Fig.1, we find the (-/2-) charge transition level of the divacancy in the gap, the ground state of $V_2(2-)$ being a triplet. Based on the position of the lowest unoccupied state, even a stable $V_2(3-)$ state appears to be plausible.

G. *The NVH center* is an important complex in N-doped CVD samples, observed in its negative charge state.[23,81] Theoretical studies[35,82,83] have established that the hydrogen atom binds to one of the carbon neighbors of the vacancy in the NV(-) center, dynamically tunneling between the three possible sites, and so exhibiting $C_{3v}$ symmetry in experiments. We have calculated the NVH defect in a static $C_{1h}$ model, as in Ref. [82]. The calculated (0/-) charge transition level agrees nevertheless very nicely with the experimental value (see Table 1).

H. *Creation of the NV(-) center*. The possibility of manipulating the optical emission and the magnetic states of the NV center makes it a very desirable defect for many applications. Obviously, control over the concentration of this particular defect and its charge state would be desirable. However, these depend on the concentration of other defects. Assuming equilibrium conditions, the calculated formation energies allow us to predict the relative concentrations in different charge states by solving the neutrality equation, considering all defects with charge $q_i$:

$$N_C \exp\left[-\frac{E_C - E_F}{kT}\right] + \sum_i |q_i| \cdot (N_{Ai} - p_{Ai}) = N_V \exp\left[-\frac{E_F - E_V}{kT}\right] + \sum_i |q_i| \cdot (N_{Di} - n_{Di}) \quad (1)$$

where

$$N_C = 2\left(\frac{2m_e^* \pi kT}{h^2}\right)^{3/2} \quad ; \quad N_V = 2\left(\frac{2m_h^* \pi kT}{h^2}\right)^{3/2} \quad (2)$$

are the effective (number)densities of states in the conduction and valence band of diamond, calculated from the density-of-states mass of the electrons, $m_e^* = 0.57\ m_0$, and the holes, $m_h^* = 0.8\ m_0$, respectively. The remaining terms in Eq.(1) are the occupancies of the acceptor and donor levels, determined by the Fermi-Dirac distribution and the degeneracy factors $g$

$$p_{Ai} = N_{Ai}\left[g_{Ai}\exp\left(\frac{E_F - E_{Ai}}{kT}\right) + 1\right]^{-1} \quad ; \quad n_{Di} = N_{Di}\left[g_{Di}\exp\left(\frac{E_{Di} - E_F}{kT}\right) + 1\right]^{-1} \quad (3)$$

The defect concentrations in Eq.(1) must be determined from the calculated energies of formation $E_{form}^{i,q}$ as

$$N_{(A,D)i} = N_{(A,D),i}^0 \exp\left(-E_{form}^{i,q}/kT\right) \quad (4)$$

for all acceptors ($A$) and donors ($D$). Here $N_i^0$ is the density of $i$ sites in the perfect lattice. We have calculated the defect formation energies with reference to the perfect 512-atom diamond supercell and the chemical potential of nitrogen in the gas phase, $\mu_N$, as

$$E_{form}^{i,q} = E^q[C_{512}:N_nV_m] - \frac{512-n-m}{512}E[C_{512}] - n\mu_N + q(E_F + E_V + \Delta V_{align}) + E_{corr}^q \quad (5)$$

where $E_{corr}^q$ and $\Delta V_{align}$ are the charge and potential alignment corrections, and $E_F$ is the Fermi-energy with respect to the valence band edge $E_V$. We have chosen $\mu_N$ to be half of the HSE06 energy of an $N_2$ molecule, $E(N_2) = -22.78$ eV, as a reference, to list the calculated formation energies in Table 4. (We also provide the formation energy of the NVH complex, using the energy of a hydrogen atom in a surface C-H bond on the 2×1-recontructed (001) surface,[84] as chemical potential for the hydrogen.[85]) Since both Eq.(1) and Eq.(5) contain the Fermi-energy, this system of equations has to be solved self-consistently.

We note that in the wide gap insulator diamond the concept of a Fermi-level – as understood in traditional semiconductors – may be of limited use at room temperature (or below),[86] but we consider here the effect of heat treatments around 1100K, where it can still be useful to understand the trends of defect formations and their charge states in diamond. Although, NV(-) centers are in practice usually not created in equilibrium processes, the study of scenarios leading to thermal equilibrium will provide insight into the formation process.

**Table 4**. HSE06 formation energies ( in eV) of the N and V related defects according to Eq.(5), with $\mu_N$ = -11.39 eV (corresponding to the energy of a nitrogen atom in the $N_2$ molecule at 0K) The formation energies of charged defects are referred to the valence band edge [$E_V$ in Eq. (5)]. The chemical potential of hydrogen was set as in Ref.[85] (see text for more details).

| Defect | $Q$ | $E_{form}^{i,q} - qE_F$ |
|---|---|---|
| N | + | 0.37 |
|   | 0 | 3.96 |
|   | - | 8.53 |
| V | 2+ | 5.72 |
|   | + | 6.15 |
|   | 0 | 7.14 |
|   | - | 9.19 |
|   | 2- | 14.05 |
| NV | + | 5.31 |
|   | 0 | 6.21 |
|   | - | 8.82 |
|   | 2- | 13.83 |
| $N_2$ | + | 2.55 |
|   | 0 | 3.92 |
| $N_2V$ | + | 4.78 |
|   | 0 | 5.41 |
|   | - | 8.64 |
| $V_2$ | + | 9.08 |
|   | 0 | 10.08 |
|   | - | 12.42 |
|   | 2- | 15.59 |
| NVH | + | 4.34 |
|   | 0 | 5.19 |
|   | - | 7.59 |
|   | 2- | 12.19 |

First, we study the equilibrium achieved after the heat treatment of nitrogen doped crystals (without prior irradiation), by assuming different nitrogen concentrations. It is known from the study of Type Ib natural diamonds that nitrogen impurities do not aggregate when the concentration of nitrogen is below 500 ppm, unless the temperature is above 2000 K. Therefore, one can exclude the formation of $N_2$ and $N_2V$ defects in a heat treatment at lower temperature. In practice, the nitrogen concentration depends on the growth conditions (temperature, pressure and nitrogen precursors present) which determine the chemical potential of nitrogen. Here we tune the value of $\mu_N$ (and with it the values in Table 4) in order to set the total concentration of nitrogen defects in the desired region between 10-500 ppm. We solved Eqs. (1-5) self-consistently under these conditions, assuming the formation of $N_s$, V, NV, and $V_2$ defects in a heat treatment at the example temperature of T=1100 K. We find that V and $V_2$ practically do not form because of their much too high formation energies. As shown in Fig.2(a), the calculated [$N_s$]/[NV] concentration ratio is constantly ~$10^3$ under these conditions, or in other words, [NV] is 0.1% of [$N_s$],. As a consequence, the Fermi-level is pinned at $E_V$ + 4.0 eV, and the vast majority of $N_s$ is neutral, and only about 0.3% will be positively and 0.2% negatively charged. Thus, about 0.1% of the $N_s$ defects donates an electron to NV defects. As a consequence, all the NV defects will be negatively charged. All-in-all, our calculations indicate that NV(-) is introduced at concentrations < 1 ppm in lightly N-doped diamond, where neutral $N_s$ (with S=1/2 electron spin) will dominate the sample. We note here that our simulation assumes infinite bulk diamond, so we do not consider surface band bending which can convert NV(-) to NV(0).[87]

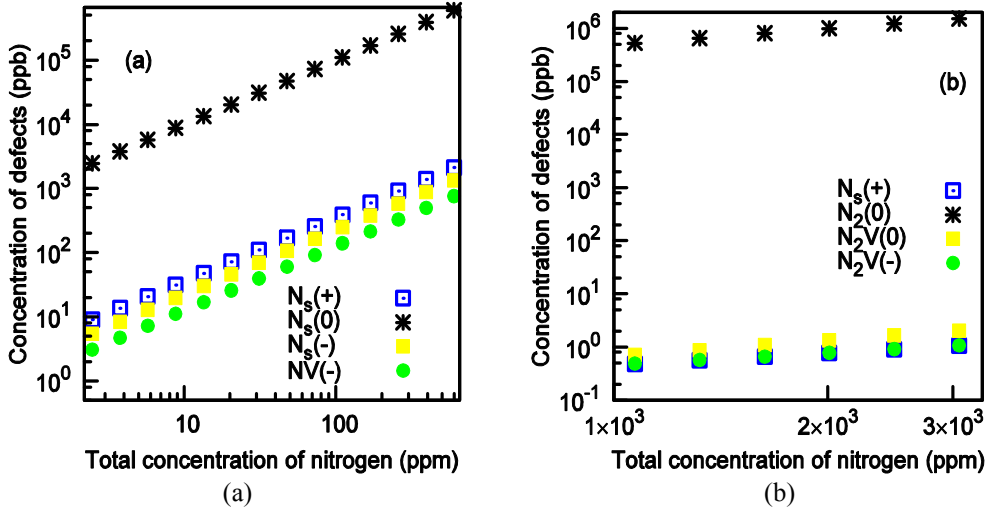

**Fig.2**. (Color online) Calculated concentrations of defects characteristic in (a) Type Ib and (b) Type IaA diamonds after annealing at 1100K. The other defects with the corresponding charge states have lower concentrations and not shown in these plots.

As next, we consider higher nitrogen contents between 1000-3000 ppm, which corresponds to Type Ia natural diamonds. In this case the average distance between N impurities is just a few lattice constants, so nitrogen impurities may aggregate even at a relatively low temperature as 1100 K, so $N_2$ and $N_2V$ may form under this condition. To simulate these conditions, we tuned $\mu_N$ to set the total concentration of nitrogen defects in the desired region and considered all the defects in all charge states as listed in Table 4, except NVH. Our simulations indicate (see Fig.2(b)) that nitrogen occurs predominantly as $N_2(0)$ while a small fraction of $N_s$ and $N_2V$ (~1 ppb) can co-exist. The NV concentration is negligible under these conditions. Since $N_2$ stays in the neutral charge state, the Fermi-level is pinned near the acceptor level of $N_2V$ at $\sim E_V + 3.2$ eV, so the neutral charge state of that defect is slightly more abundant than the negative one.

Synthetic diamonds can also be grown by chemical vapor deposition (CVD), with substrate temperatures around 1100K. Here the formation of NV is influenced by hydrogen impurities which enter the crystal in the CVD process. According to recent experiments, NVH defects (see III.G) form in a ratio of 0.01-0.02 to the incorporated $N_s$, when the concentration of $N_s$ is about 0.5-1.2 ppm.[23,88,89] The concentration of NV is below the detection limit of 0.1 ppb in these samples, which means [NV]/[$N_s$] < 0.1%. According to the calculated formation energies (Table 4), the NVH complex has about 1 eV lower formation energy than that of NV. This result explains why the NVH defect can outcompete the NV defect in CVD diamond. The NVH complex is stable against annealing up to 1600 °C.[23,90] Above that temperature NV defects can already diffuse, too, thus NVH defects cannot be converted to NV by thermal annealing. So, the NV concentration in CVD samples is again insufficient for practical applications.

In practice, the concentration of NV(-) centers can be increased by irradiation and subsequent annealing. The irradiation creates Frenkel-pairs (and other damage) in the diamond lattice. Annealing leads to recombination, but some Frenkel-pairs may split to produce isolated vacancies and self-interstitials with concentrations much above that of thermal equilibrium. The self-interstitials are mobile even at room temperature and will aggregate to the surface or grain boundaries, or form platelet-like defects. In the mean time they can assist nitrogen diffusion and aggregation, too. Subsequent to irradiation, a heat treatment has to be applied to anneal out luminescence-quenching parasitic defects. This is usually done slightly above 600 °C, where neutral vacancies become mobile. It is usually assumed that NV centers are formed during this heat treatment when vacancies get trapped at $N_s$ defects. However, vacancies may also get trapped at existing $N_2$ defects, or can form divacancies. The post-irradiation annealing can be regarded as a quasi equilibrium process, and an insight into the creation of NV(-) centers can be gained by close inspection of the formation energies and occupation levels of the considered defects. First of all, one can assume that the initial concentration of isolated $N_s$ defects is high enough to pin the Fermi-level initially above midgap. In order to have mobile, i.e., neutral vacancies after the irradiation, the Fermi-level must be lowered drastically, below the single acceptor level of V (at about $E_V + 2.0$ eV). Thus, if NV defects are to be created by irradiation and annealing, the vacancy concentration should be in excess of the $N_s$ concentration ([V] > [$N_s$]), even *after* the trivial recombination with interstitials. Then, two basic reactions can occur:

$$V(0) + V(0) \rightarrow V_2(0) + 4.2 \text{ eV} \qquad (6)$$

$$V(0) + N_s(+) \rightarrow NV(0) + h + 3.3 \text{ eV} \qquad (7)$$

where *h* is a hole with energy corresponding to the given Fermi-level position. Both reactions are strongly exothermic, as can be derived from the data in Table 4.[91] Since [V] > [$N_s$] and Eq.(6) provides a higher energy gain than Eq.(7), the majority of the vacancies will form divacancies and only a small fraction creates NV defects. In fact, since the formation of $V_2$ is about 0.9 eV more favorable than that of NV, the equilibrium concentration of $V_2$ will be several orders of magnitude larger than that of NV, even at relatively high temperatures (at 1100K by a factor of $2 \cdot 10^4$). This implies that the concentration of NV defects, arising through the reaction in Eq.(7), will not be significantly higher than without irradiation. We also note that the generally assumed process of creating NV defects by vacancy diffusion would be self-limiting, anyhow. As isolated vacancies start to form $V_2$ and NV defects, the Fermi-level shifts up because both $V_2$ and NV are deeper acceptors than V (c.f. Figure 1). As a result, the remaining isolated vacancies will become negatively charged and immobilized. So increasing the vacancy concentration cannot really help to increase [NV].

The observed increase in [NV] can, therefore, be explained only by assuming that NV defects dominantly form during irradiation, not during the annealing. Our results support this assumption. With the data of Table 4, the creation of a vacancy near to $N_s$ requires an energy of

$$N_s(0) + 2.26 \text{ eV} \rightarrow NV(0) + C \qquad (8)$$

while that of a vacancy in a perfect part of the crystal needs

$$\text{perfect lattice } + 7.14 \text{ eV} \rightarrow V(0) + C \qquad (9)$$

where C is a carbon atom in the perfect diamond lattice. The reason for the difference is that to remove the C atom opposite to $N_s$ requires to break only three strong C-C bonds (see III.A), whereas four such bonds have to be broken in the perfect diamond lattice to form an isolated vacancy. Such a big energy difference should lead to a strong preference for NV creation even in the non-equilibrium process of irradiation, explaining most of the arising NV concentration. We conclude, therefore, that the dominant part of the NV concentration is created directly by the irradiation.

According to our simplified model, the dominant point defects in N-doped, irradiated and annealed diamond samples are $N_s$, NV and $V_2$. The charge state of the NV defect will depend on the relative concentrations of the $N_s$ donors and the $V_2$ acceptors. To study the chances for creating negatively charged NV-centers, we have tuned the formation energies of these three defects to obtain a total nitrogen concentration of 386 ppm and an $N_s$-to-NV conversion factor of 1.4% (i.e., within the range of experimental observations between 0.5-2.5%), at various $[V_2]/[N_s]$ ratios. Fig. 3 shows how the $[NV(-)]/[NV(0)]$ ratio depends on $[V_2]/[N_s]$.

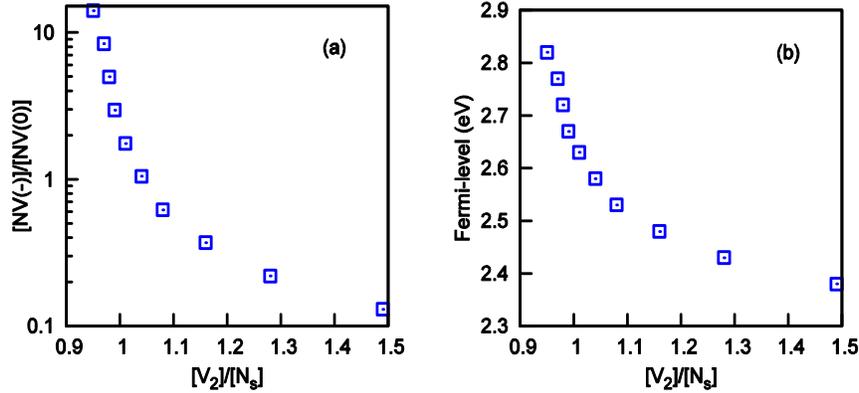

**Fig.3**. The calculated (a) concentration ratio $[NV(-)] / [NV(0)]$ and (b) the corresponding Fermi-level position (with respect to the valence band edge) as a function of the ratio of $[V_2] / [N_s]$ at T=1100 K. The total concentration of nitrogen is set to ~386 ppm while the $N_s$-to-NV conversion factor to 1.4%.

If divacancies dominate, i.e., $[V_2]/[N_s] > 1$, then the Fermi-level will be pinned near the single acceptor level of $V_2$ at $E_V + 2.3$ eV. Since the first acceptor level of NV is at $E_V + 2.7$ eV, our simulation results in an $[NV(-)]/[NV(0)]$ concentration ratio of ~0.1. Therefore, for $[V_2] > [N_s]$ the neutral NV would dominate. Reducing $[V_2]$ will shift the Fermi-level towards the acceptor level of NV and, as soon as $[V_2]/[N_s] < 1$, the negative charge state of NV becomes dominant. Our simulation demonstrates (see Fig. 3) that the charge state of NV is very sensitive to the concentration of $V_2$ in this range. Changing the concentration of $V_2$ by less than a factor of two, can change the [NV(-)]/[NV(0)] ratio by a factor of ~100.

These results show that the post-irradiation annealing not only does not contribute significantly to the NV production but, by creating divacancies, may prevent the achievement of negatively charged NV defects. Of course, the annealing is unavoidable, but our analysis indicates that its temperature should be chosen in the range where also $V_2$ becomes mobile, while NV yet does not. This is possible as the TH5 center associated with $V_2$ starts to anneal out from 800 °C,[78,92] where NV is not yet mobile. (We emphasize here that annealing out $V_2(0)$ defects can raise the Fermi-level, which will change the charge state of the residual $V_2$ from neutral to negative. Thus, the $V_2(-)$ signals should be also monitored beside $V_2(0)$ to determine the concentration of the remaining divacancies at elevated temperatures.)

The annihilation of the divacancies may occur by out-diffusion, but also via recombination at interstital clusters, or by the formation of vacancy aggregates which are also electrically active.[93,94,95,96] However, it appears likely that the vacancy aggregates are acceptor defects, with a charge transition level at about $E_V + 3.5$ eV.[23] This is well above the (0/-) level of NV, so they can donate electrons to turn NV(0) to NV(-). Thus, elimination of $V_2$ can stabilize the charge state of NV(-).

Our analysis is in line with the observed higher efficiency of NV(-) creation in annealing irradiated diamonds at higher than usual temperatures (1100-1200 °C).[31] The annihilation of $V_2$ is important even when [NV(-)]/[NV(0)] > 1 happens to be the case after irradiation and annealing, because $V_2$ will be negatively charged under this condition, and can compromise the photo-stability of NV(-). Indeed, high-temperature post-annealing treatments could help stabilizing the charge state of NV(-).[22] Our results highlight the need of careful characterization of irradiated and annealed diamond samples, particularly, focusing on divacancy or larger vacancy aggregates.[97]

IV. SUMMARY AND CONCLUSIONS

We calculated the charge transition levels, excitation energies, barrier energy for migration and reaction energies of basic vacancy and nitrogen related defects by the HSE06 supercell plane wave method. We have reproduced the known experimental data regarding electronic transitions, substantially improving over previous (standard) DFT calculations. In particular, without any *a posteriori* correction, our HSE06 calculation reproduces *all* experimentally observed charge transition levels and internal transitions within 0.2 eV. In contrast to standard DFT, we find that the relaxation of the atoms around a neutral vacancy (larger in the radial than in the axial direction of $D_{2d}$) splits the doubly occupied $t_2$ single-particle state into a *lower lying e* and a higher lying $b_2$ state. A spin-triplet occupation of the *e* state realizes the $^3T_1$ excited many-particle state which is know to lie 0.1 eV above the $^1E$ ground state. Taking that into account, the position of the (+/0) charge transition level (the only one known experimentally) is reproduced within 0.1 eV. Unlike standard DFT, HSE06 correctly finds the symmetry of the neutral and the spin state of the negative divacancy. Our calculations also provide the first explanation for the observed luminescence of $N_2V$, reproducing the observed transition energies within 0.2 eV in both the singlet and triplet recombination channels. The proven accuracy of the method has allowed us to predict missing data on the charge transitions of all the investigated defects ($N_s$, V, NV, NVH, $N_2$, $N_2V$ and $V_2$), which are crucial to establish the charge state of different defects. Our results also comply with the experimental finding on the migration of isolated vacancy, namely, that only its neutral form is mobile while it is immobile in its negative charge state.

By assuming quasi-equilibrium conditions, we found that the NV center may be created in lightly nitrogen-doped diamond ([N] < 500 ppm) in small concentration, whereas the formation of $N_2V$ defects is much more likely for high concentrations ([N] > 1000 ppm). We also investigated the basic reaction for the formation of NV centers in irradiated and annealed samples. The key findings are that

- *i)* Irradiation is more likely to directly create NV defects than vacancies.
- *ii)* In post-irradiation annealing much more divacancies are formed than NV defects, and only short range diffusion of vacancies towards proximate substitutional nitrogen defects can increase the concentration of NV centers.
- *iii)* Since the divacancy is a deeper acceptor than NV, the created NV defects will dominantly be in the neutral charge state, unless the concentration of divacancies is sufficiently decreased by annealing above ~1100 K.
- *iv)* Remaining divacancies may influence the photo stability of NV(-) centers.

ACKNOWLEDGEMENT

AG gracefully acknowledges the support from FP7 grants DIAMANT and DIADEMS of the EU Commission. The support of the supercomputer center HLRN (Grant. No. 0011) is very much appreciated. Fruitful discussions with J. Wrachtrup are appreciated.